\documentclass[prc,twocolumn,superscriptaddress,nofootinbib]{revtex4-1}
\usepackage{newtxtext}
\usepackage[varvw,bigdelims]{newtxmath}
\usepackage[colorlinks=true,allcolors=blue]{hyperref}
\usepackage{graphicx}
\usepackage{bm}
\usepackage{amsmath}
\usepackage{xcolor}

\begin{document}
\title{\boldmath 
$\Lambda(1670)$ production in the $\psi(3686) \to  \Lambda \bar \Lambda \eta$ reaction
} 

\author{Natsumi Ikeno}
\email{ikeno@maritime.kobe-u.ac.jp}
\affiliation{Graduate School of Maritime Sciences, Kobe University, Kobe 658-0022, Japan}

\author{Eulogio Oset}
\email{Eulogio.Oset@ific.uv.es}
\affiliation{Departamento de F\'{i}sica Teórica and IFIC, Centro Mixto Universidad de Valencia-CSIC, Institutos de Investigaci\'{o}n de Paterna, Aptdo. 22085, E-46071 Valencia, Spain}

\begin{abstract}
We perform a calculation of the invariant mass distributions in the $\psi(3686) \to  \Lambda \bar \Lambda \eta$ reaction, where a neat peak for the excitation of the $\Lambda(1670)$ and $\bar \Lambda(1670)$ in the $\eta \Lambda$ and $\eta \bar \Lambda$ mass distributions, respectively, is observed. Our approach uses the fact that the 
$\psi(3686)$, a $c \bar c$ state, is a singlet of SU(3) in the $u,d,s$ quarks and constructs the two flavor structures allowed with a pseudoscalar meson, a baryon and an antibaryon. The resonance peaks come from the final state interaction of meson baryon pairs, which generate the $ \Lambda(1670)$ in our approach. With a reasonable relative weight of the two flavor structures, the only free parameter of the theory, we are able to get the three mass distributions in good agreement with experiment, giving extra support to the molecular structure of the  $ \Lambda(1670)$ resonance.
The agreement with data improves with an extra resonant contribution with mass around 2200 MeV, possibly accounting for some resonances. 
\end{abstract}

\date{\today}

\maketitle
\section{Introduction}
The $\Lambda(1670)$ has been known for some time from the study of $\bar K p $ collisions~\cite{Gopal:1976gs}, but the advent of new facilities has brought large sets of data in new reactions where it is also observed with great precision. This is the case of the Belle reaction $\Lambda_c \to \Lambda \pi^+ \eta$~\cite{Belle:2020xku}, where a clear signal is seen for the $\Lambda(1670)$  in the $\eta \Lambda$ invariant mass distribution. The resonance is also very clearly seen in the $\eta \Lambda$ mass spectrum in the $J/\psi \to \Lambda \bar \Sigma \eta$ decay~\cite{BESIII:2026vrm} from the BESIII collaboration. In another Belle work~\cite{Belle:2022cbs}, a different reaction is investigated, the $\Lambda_c \to  \pi^+ p K^-$ decay, and a clear peak is also observed at the $p K^- $ threshold, which also reflects the $\Lambda(1670)$ excitation.  The high resolution and statistics of this latter reaction improves appreciably earlier results of LHCb, where the peak was also observed~\cite{LHCb:2022sck}.

The $\Lambda(1670)$ plays an important role in chiral dynamics of the meson baryon interaction. It was obtained from the interaction of the SU(3) octet of pseudoscalar mesons with the octet of baryons \cite{Oset:2001cn,Garcia-Recio:2002yxy,Kolomeitsev:2003kt,Oller:2006jw} using chiral Lagrangians as a source 
of the potential, which is later unitarized leading to the chiral unitary approach~\cite{Oset:1997it}. One interesting thing was observed in the work of Ref.~\cite{Jido:2003cb}: 
The chiral Lagrangian used as the source of the interaction, which leads to the potential entering the Schr\"{o}dinger (or Bethe–Salpeter) equation, is SU(3) symmetric. However, the unitarization procedure, which accounts for the propagation of intermediate meson–baryon coupled channels, breaks SU(3) symmetry when different masses of  the particles belonging to the same SU(3) multiplet (for instance, pions and kaons) are used. 
In Ref.~\cite{Jido:2003cb} an exercise was done, rendering all the masses equal for the particles belonging to the same SU(3) multiplet, in which the solution was SU(3) symmetric, and the interaction generated dynamically two bound states, one for an octet and another one for a singlet of the meson-baryon system. Then, the SU(3) symmetry was gradually broken and the octet produced two branches of isospin $I=0$ and two of $I=1$. 
One $I=0$ branch leads to one $\Lambda(1405)$ state 
($\Lambda(1420))$, 
which, together with the branch of the singlet, leads to the two $\Lambda(1405)$ states, 
now officially included  in the PDG~\cite{ParticleDataGroup:2024cfk}, while the other $I=0$ branch moved to produce the $\Lambda(1670)$. It is very interesting to observe that the present $\Lambda(1420)$ and the $\Lambda(1670)$ have the same dynamical origin and they are the same state in the strict SU(3) limit, and the reason for the 250 MeV difference stems from the breaking of SU(3) symmetry due to the propagation of the intermediate meson baryon components in the unitarization procedure (Bethe Salpeter series of terms in the scattering amplitudes) when the physical masses of the particles are considered. This is a clear example of how SU(3) symmetry can be broken in nature, even if the original Lagrangians that provide the interaction are SU(3) symmetric. In other words, the unitarization of the amplitudes, or equivalently, the final state interaction after a primary production process, is essential to properly interpret hadronic reaction mechanisms.  

The reactions mentioned above also bear information on the nature of the  $\Lambda(1670)$, which is not universally accepted as a molecular state. For instance, a 3 quark nature is assumed in Refs.~\cite{CrystalBall:2001uhc,Zhong:2008km}. One way to find support for the nature of a resonance is to make predictions for their production in different reactions. In this sense, it was suggested in the theoretical work of Ref.~\cite{Miyahara:2015cja} to study the production of the $\Lambda(1670)$ in the decay of $\Lambda_c \to \pi^+ MB$, where $MB$ are a pseudoscalar meson and a baryon of the SU(3) octets. It was found that the $\Lambda(1405)$ was seen in the $MB=\pi \Sigma$ channel, while the $\Lambda(1670)$ was clearly seen in the $\eta \Lambda$ channel. These results were corroborated by the calculations of Ref.~\cite{Xie:2016evi} in the study of the $\Lambda^+_c \to \pi^+ \eta \Lambda$ decay. The suggestion led to the Belle experiment  $\Lambda_c \to \Lambda \pi^+ \eta$~\cite{Belle:2020xku}, where indeed a very clear signal was seen in the $\eta \Lambda$ invariant mass distribution. The reaction has been posteriorly studied in Refs.~\cite{Wang:2022nac,Duan:2024czu,Lyu:2024qgc,Lyu:2026ack}, where an agreement with the different mass distributions is found, using the dynamics in which the $\Lambda(1670)$ is dynamically generated from the same coupled channels as the  two $\Lambda(1405)$.

  The $\Lambda(1670)$ is also seen in a peculiar reaction from the LHCb and Belle collaborations, the $\Lambda_c^+ \to p K^- \pi^+$,  where there is a cusp like peak at the $\eta \Lambda$ threshold which is associated to the $\Lambda(1670)$~\cite{LHCb:2022sck,Belle:2022cbs}. The reaction was studied from the molecular perspective for the $\Lambda(1670)$ resonance in Refs.~\cite{Duan:2024okk,Zhang:2024jby}. 
A peak was observed in  the $K^- \pi^+$ invariant-mass distribution, which the theoretical framework could  associate  with the $\Lambda(1670)$ excitation. At the same time, the $\eta \Lambda$ scattering length and effective range were determined, yielding values that are consistent with, and improve upon, those obtained from the $K^- p \to \eta \Lambda$ reaction~\cite{CrystalBall:2001uhc}.

  The interesting thing about these reactions is that the $\Lambda(1670)$ resonance shows up with different line shapes, due to interference with other terms, in particular with the tree level contributions of the channels in which the resonance is observed. But some times the tree level contribution does not appear and one can see the resonance more clearly. This is, for instance, the case of the $J/\psi \to \Lambda \bar \Sigma \eta$ decay~\cite{BESIII:2026vrm}, which is isospin forbidden and the tree level is zero. The reaction has been studied theoretically in Ref.~\cite{Dai:2026zqn}. Isospin is broken through loops that do not cancel due to different physical masses of particles in the same isospin multiplets, but these loops are precisely what give rise to the dynamically generated resonances, and hence this is particularly valuable to find support for the molecular picture of resonances, as is the case in Ref.~\cite{Dai:2026zqn}, which finds support for the molecular structure of the $\Lambda(1670)$.

   In the present case, we study a related reaction, the $\psi(3868) \to \Lambda \bar \Lambda \eta$,  measured by the BESIII collaboration~\cite{BESIII:2022cxi}. In this case, the reaction is isospin allowed and one finds indeed a line shape quite different to the one found for the  $J/\psi \to \Lambda \bar \Sigma \eta$ decay, which makes the theoretical explanation challenging. 
We follow the same procedure for this reaction and employ the same theoretical description of the $\Lambda(1670)$ as in Ref.~\cite{Dai:2026zqn}, comparing the resulting predictions with the experimental data.
With basically one free parameter, we show that we can obtain a good agreement with the experimental mass distributions in all the energy range, and show the role played by the $\Lambda(1670)$ in the reaction.
However, we see that a very good agreement with data demands 
the contribution of an extra resonance term with mass around 2200 MeV, possibly coming from more than one resonance.

\section{Formalism}\label{sec:formalism}
The reaction that we study is $\psi(3686) \to \Lambda \bar \Lambda \eta$. We first look at the flavor structure of the $\psi \to B \bar{B} P$, with $B$ a baryon of the SU(3) octet and $P$ a pseudoscalar meson.
Since $\psi(3686)$ is a $c\bar{c}$ state, a scalar object in SU(3), then we can construct SU(3) invariant combinations of $B \bar{B} P$. This is accomplished by taking traces of $B \bar{B} P$ using the SU(3) matrices for $B$, $\bar{B}$, and $P$,
\begin{equation}
P=
\begin{pmatrix}
\frac{1}{\sqrt{2}}\pi^0 + \frac{1}{\sqrt{3}} \eta & \pi^+ & K^+ \\[2mm]
\pi^- & -\frac{1}{\sqrt{2}} \pi^0 + \frac{1}{\sqrt{3}} \eta & K^0 \\[2mm]
K^- & \bar{K}^0 & -\frac{1}{\sqrt{3}} \eta
\end{pmatrix},
\label{eq:Pmatrix}
\end{equation}

\begin{equation}
B=
\begin{pmatrix}
\frac{1}{\sqrt{2}}\Sigma^0 + \frac{1}{\sqrt{6}} \Lambda & \Sigma^+ & p\\[2mm]
\Sigma^- & -\frac{1}{\sqrt{2}} \Sigma^0 + \frac{1}{\sqrt{6}} \Lambda & n \\[2mm]
\Xi^- & \Xi^0 & -\frac{2}{\sqrt{6}} \Lambda
\end{pmatrix},
\label{eq:Bmatrix}
\end{equation}

\begin{equation}
\bar{B}=
\begin{pmatrix}
\frac{1}{\sqrt{2}}\bar{\Sigma}^0 + \frac{1}{\sqrt{6}} \bar{\Lambda} & \bar{\Sigma}^+ & \bar{\Xi}^+ \\[2mm]
\bar{\Sigma}^- & -\frac{1}{\sqrt{2}} \bar{\Sigma}^0 + \frac{1}{\sqrt{6}} \bar{\Lambda }&  \bar{\Xi}^0\\[2mm]
\bar{p} & \bar{n} & -\frac{2}{\sqrt{6}} \bar{\Lambda}
\end{pmatrix},
\label{eq:Bbarmatrix}
\end{equation}
where we have omitted the $\eta^\prime$ field which does not play a role here.

We can have the traces 
$\langle \bar{B} B P \rangle$, $\langle \bar{B} P B \rangle$,  $\langle \bar{B} B\rangle \langle P \rangle$, $\langle \bar{B} P\rangle \langle B \rangle$, $\langle B P\rangle \langle \bar{B} \rangle$, and $\langle  \bar{B} \rangle \langle  B \rangle \langle  P \rangle$. 
Since $\langle B \rangle = \langle \bar B \rangle =0 $, only $\langle \bar{B} B P \rangle$, $\langle \bar{B} P B \rangle$,  $\langle \bar{B} B\rangle \langle P \rangle$ survive.
Actually $\langle P \rangle$ is also zero if we take the octet of pseudoscalars.
The $\langle P \rangle$ that we have does not vanish because we also have a mixture with the singlet of SU(3) to implement the $\eta$--$\eta'$ mixing of Ref.~\cite{Bramon:1992kr}. In any case, according to large $N_c$ arguments~\cite{Manohar:1998xv,Abreu:2023yvf,Dai:2026zqn}, terms with more than one trace are suppressed versus those with a single trace. Then we are left with the two flavor structures, $\langle \bar{B} B P \rangle$ and $\langle \bar{B} P B \rangle$.
These structures are then equivalent to the ones used in Ref.~\cite{He:2026mkf},
\begin{equation}
\mathcal{L}_\psi =
\tilde{D}\,\langle \bar{B} \gamma_\mu \gamma_5 \{ P, B \} \rangle\, \psi^\mu
+
\tilde{F}\,\langle \bar{B} \gamma_\mu \gamma_5 [ P, B ] \rangle\, \psi^\mu.
\label{eq:Lagrangian}
\end{equation}
Since we want to have a $\Lambda \bar{\Lambda} \eta$ at the end, and we want to consider the $BP$ interaction, we single out the terms that have $\Lambda$ or $\bar{\Lambda}$ and any pairing of $\bar{B}P$ or $BP$, respectively, and we find:\\

(1)~$\langle \bar{B} P B \rangle$ structure:
\begin{align}
&\frac{\bar{\Lambda}}{\sqrt{6}} \Big(
\pi^0 \Sigma^0  +  \pi^+ \Sigma^- + \pi^- \Sigma^+ +   K^+ \Xi^- +  K^0 \Xi^0
- 2 K^- p - 2 \bar{K}^0 n \notag \\
&\qquad
- \frac{2}{\sqrt{3}\sqrt{6}}\, \eta \Lambda
\Big),
\label{eq:BPB_Lambar}\\
&\frac{\Lambda}{\sqrt{6}} \Big(
\bar{\Sigma}^0 \pi^0 + \bar{\Sigma}^+ \pi^- + \bar{\Sigma}^- \pi^+ + \bar{\Xi}^+ K^- + \bar{\Xi}^0 \bar{K}^0 
- 2\bar{p} K^+ - 2\bar{n} K^0 \notag \\
&\qquad
- \frac{2}{\sqrt{3}\sqrt{6}}\,\bar{\Lambda}\eta
\Big).
\label{eq:BPB_Lam}
\end{align}

(2)~$\langle \bar{B} B P\rangle$ structure:
\begin{align}
&\frac{\bar{\Lambda}}{\sqrt{6}} \Big(
\Sigma^0 \pi^0  +  \Sigma^- \pi^+ + \Sigma^+ \pi^- + p K^- + n \bar{K}^0  
- 2  \Xi^- K^+ -  2 \Xi^0 K^0
\notag \\
&\qquad
- \frac{2}{\sqrt{3}\sqrt{6}}\, \Lambda \eta
\Big),
\label{eq:BBP_Lambar}\\
&\frac{\Lambda}{\sqrt{6}} \Big(
\pi^0 \bar{\Sigma}^0 + \pi^+ \bar{\Sigma}^- + \pi^- \bar{\Sigma}^+ 
+ K^+ \bar{p}  + K^0 \bar{n} 
- 2 K^- \bar{\Xi}^+ - 2 \bar{K}^0 \bar{\Xi}^0 
 \notag \\
&\qquad
- \frac{2}{\sqrt{3}\sqrt{6}}\, \eta \bar{\Lambda}
\Big).
\label{eq:BBP_Lam}
\end{align}
Note that the terms with $\Lambda \bar{\Lambda}$ in Eqs.~\eqref{eq:BPB_Lambar} and \eqref{eq:BPB_Lam} are the same, and the same holds for Eqs.~\eqref{eq:BBP_Lambar} and \eqref{eq:BBP_Lam}, so it should not be counted twice. Next, we give weight $\tilde{A}$ to the terms in Eqs.~\eqref{eq:BPB_Lambar} and \eqref{eq:BPB_Lam}, 
and weight $\tilde{B}$ to the terms in Eqs.~\eqref{eq:BBP_Lambar} and \eqref{eq:BBP_Lam}, and we finish having the flavor structure.

\begin{align}
\psi  \to  &
\frac{\bar \Lambda}{\sqrt{6}} \Big\{
(\tilde{A}-2\tilde{B}) \left( K^+ \Xi^- + K^0 \Xi^0 \right)  \notag \\
& \quad
+ (-2\tilde{A}+\tilde{B}) \left( K^- p + \bar{K}^0 n \right)
\Big\} \notag \\
+& \frac{\Lambda}{\sqrt{6}} \Big\{
(\tilde{A}-2\tilde{B}) \left( K^- \bar{\Xi}^+ + \bar{K}^0 \bar{\Xi}^0 \right) 
\notag \\
& \quad
+ (-2\tilde{A}+\tilde{B}) \left( K^+ \bar{p} + K^0 \bar{n} \right)
\Big\} \notag \\
-& 2(\tilde{A}+\tilde{B}) \frac{1 }{6\sqrt{3}}\, \eta \Lambda \bar{\Lambda},
\label{eq:psi_LLbeta}
\end{align}
where the $\pi \Sigma$ terms have been eliminated, because in Ref.~\cite{Oset:2001cn}, we found that the couplings of the $\pi \Sigma$
channel to the $\Lambda(1670)$ are extremely small, and we neglect them.

By taking into account the isospin multiplets
$(K^+, K^0)$, $(\bar{K}^0, -K^-)$, $(\Xi^0, -\Xi^-)$, $(p,n)$,
we can write
\begin{align}
\psi \to & \frac{\bar \Lambda}{\sqrt{6}} \Big\{
(2\tilde{B}-\tilde{A}) \sqrt{2}\, |K \Xi, I=0\rangle
+ (\tilde{B}-2\tilde{A}) \sqrt{2}\, |\bar{K} N, I=0\rangle
\Big\} \notag\\
+& \frac{\Lambda}{\sqrt{6}} \, \{ T_{\text{bar}}  \}
\notag\\
-&\frac{1}{3\sqrt{3}} (\tilde{A}+\tilde{B})\, \eta \Lambda \bar{\Lambda},
\label{eq:Tiso}
\end{align}
 where $\{T_{\text{bar}}\}$ stands for the terms of the bracket $\{\}$ of Eq.~\eqref{eq:Tiso} obtained by changing the particles into their  antiparticles, without the need of classification into isospin states.

\begin{figure*}[!htb]
\begin{center}
\includegraphics[width=0.6\linewidth]{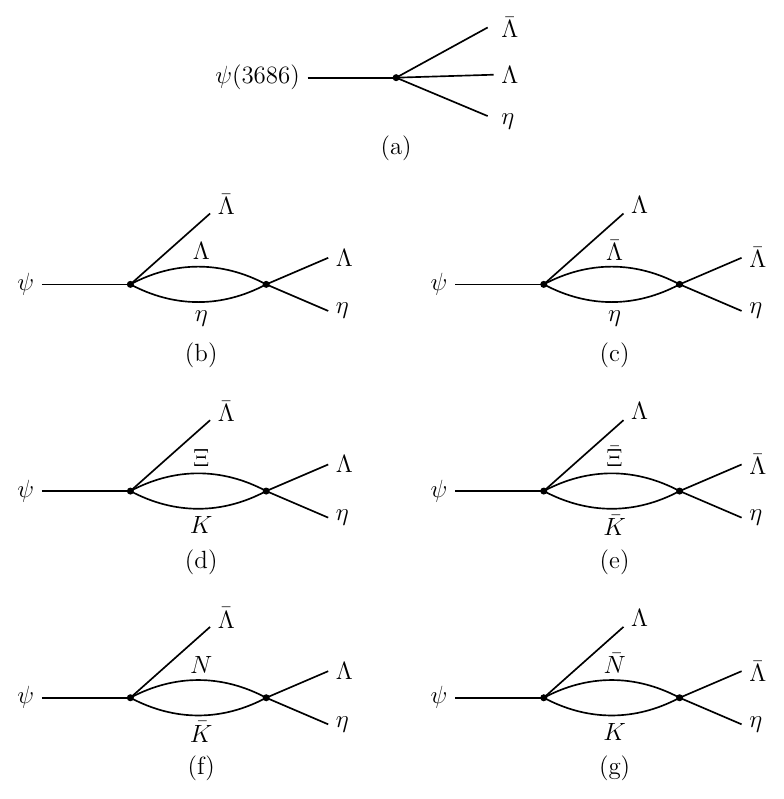}
\caption{
Terms entering the production of $\Lambda \bar{\Lambda} \eta$
in the $\psi$ decay. The first diagram (a) stands for the tree level,
and the other ones (b)--(g) for rescattering that produce the $\Lambda(1670)$.}
\label{fig:diagram}
\end{center}
\end{figure*}

In order to produce $\Lambda \bar{\Lambda} \eta$ in the final state, we can obtain it not only from the direct term with $\eta \Lambda \bar{\Lambda}$ in Eq.~\eqref{eq:Tiso} (tree level), but also from rescattering processes,
such as $\eta \Lambda \to \eta \Lambda$, $\eta \bar \Lambda \to \eta \bar \Lambda$, 
$ K \Xi \to \eta \Lambda$, $\bar{K} \bar \Xi \to \eta \bar \Lambda$,
which we depict in Fig.~\ref{fig:diagram}.

We next write the general amplitude that we have for this process. The basic structure, according to Eq.~\eqref{eq:Lagrangian}, is
\begin{equation}
t \equiv - \overline{u}_{B}(\bm{p})\, \gamma_\mu \gamma_5\, v_{\bar{B}}(\bm{p}^\prime)\, \epsilon^\mu 
 \equiv  \overline{u}_{B}(\bm{p})\, \gamma^i \gamma_5\, v_{\bar{B}}(\bm{p}^\prime)\, \epsilon^i ,
\label{eq:t_general}
\end{equation}
where we consider that $\epsilon^0=0$ in the $\psi$ rest frame.
We keep terms in the spinors up to linear terms in the momenta, where 
\begin{equation}
u_{B}(\bm{p}) = 
\begin{pmatrix}
\chi_r \\
\dfrac{\bm{\sigma}\cdot \bm{p}}{2m}\chi_r
\end{pmatrix},
\qquad
v_{\bar{B}}(\bm{p}^\prime) = 
\begin{pmatrix}
\dfrac{\bm{\sigma}\cdot \bm{p}^\prime}{2m^\prime} \chi^\prime_r \\
\chi^\prime_r
\end{pmatrix},
\label{eq:uv_spinor}
\end{equation}
and then get the structure of the primary production amplitude
\begin{equation}
\overline{u}(\bm{p}) \gamma^i \gamma_5 v(\bm{p}^\prime)\, 
\;\to\;
\frac{ p^i}{2m} + \frac{p^{\prime i}}{2m^\prime}
+ i \epsilon_{ijk} \sigma_k \left( \frac{p^{\prime j}}{2m^\prime} - \frac{p^j}{2m} \right),
\label{eq:t_new}
\end{equation}
with $m$, $m^\prime$ the masses of $\Lambda$, $\bar \Lambda$.
Then, in the rescattering terms, one of the momenta corresponds to the spectator baryon, while the other corresponds to the baryon in the loop. Since the scattering amplitudes $BP \to \eta \Lambda$, $\bar B P \to \eta \bar \Lambda$ are in $s$-wave, the terms linear in the momenta of internal baryons vanish in the loop integration and taking this into account, we obtain the following amplitude for all the terms in Fig.~\ref{fig:diagram}.
\begin{align}
t =& \Bigg\{
D \left( \frac{p_{\Lambda}^i}{2M_{\Lambda}} - i \epsilon_{ijk} \sigma_k \frac{p_{\Lambda}^j}{2M_{\Lambda}} \right) 
 + D^\prime \left( \frac{p_{\bar{\Lambda}}^i}{2M_{\Lambda}}
 + i \epsilon_{ijk} \sigma_k \frac{p_{\bar{\Lambda}}^j}{2M_{\Lambda}} \right) 
\Bigg\}\notag \\
& \cdot \epsilon^i
\label{eq:t_DDEE}
\end{align}
where
\begin{align}
D = & -\frac{1}{3\sqrt{3}} (\tilde{A}+\tilde{B})  \notag\\
& -\frac{1}{3\sqrt{3}}(\tilde{A}+\tilde{B})\,
 G_{\eta\Lambda}(M_{\text{inv}}(\eta \bar{\Lambda}) )\,
 t_{\eta\Lambda,\eta\Lambda}(M_{\text{inv}}(\eta \bar{\Lambda}) ) \notag\\
&+ \frac{1}{\sqrt{3}} (2\tilde{B}-\tilde{A}) \,
G_{K \Xi}(M_{\text{inv}}(\eta \bar{\Lambda}))\,
t_{K \Xi \to \eta\Lambda}(M_{\text{inv}}(\eta \bar{\Lambda})) \notag \\
&+ \frac{1}{\sqrt{3}}(\tilde{B}-2\tilde{A}) \,
G_{\bar K N}(M_{\text{inv}} (\eta \bar{\Lambda}))\,
t_{\bar K N \to \eta\Lambda}(M_{\text{inv}} (\eta \bar{\Lambda})), 
\label{eq:D}\\
D^\prime = & -\frac{1}{3\sqrt{3}} (\tilde{A}+\tilde{B})  \notag\\
& -\frac{1}{3\sqrt{3}}(\tilde{A}+\tilde{B})\,
 G_{\eta\Lambda}(M_{\text{inv}}(\eta \Lambda) )\,
 t_{\eta\Lambda,\eta\Lambda}(M_{\text{inv}}(\eta \Lambda) ) \notag\\
&+ \frac{1}{\sqrt{3}} (2\tilde{B}-\tilde{A}) \,
G_{K \Xi}(M_{\text{inv}}(\eta \Lambda))\,
t_{K \Xi \to \eta\Lambda}(M_{\text{inv}}(\eta \Lambda)) \notag \\
&+ \frac{1}{\sqrt{3}}(\tilde{B}-2\tilde{A}) \,
G_{\bar K N}(M_{\text{inv}} (\eta \Lambda))\,
t_{\bar K N \to \eta\Lambda}(M_{\text{inv}} (\eta \Lambda)), 
\label{eq:Dp} 
\end{align}
In Eqs.~\eqref{eq:t_DDEE}--\eqref{eq:Dp}, we have taken into account that
$t_{BP, B^\prime P^\prime} = t_{\bar{B} \bar{P}, \bar{B}^\prime \bar{P}^\prime}$.
By summing and averaging over spins of the particles, we obtain
\begin{align}
\overline{\sum}\sum |t|^2
= &
\frac{1}{6 m_\Lambda^2}
\Big[ 3|D|^2 \,\bm{p}_{\Lambda}^{\,2}
 + 3|D^\prime|^2 \,\bm{p}_{\bar{\Lambda}}^{\,2}
\nonumber \\
&\qquad
-2\,\mathrm{Re}(D D^{\prime *})
\bm{p}_{\Lambda}\cdot \bm{p}_{\bar{\Lambda}}
\Big],
\label{eq:t_final}
\end{align}
which depends on the invariant masses. Actually
\begin{equation}
\bm{p}_{\Lambda}\cdot \bm{p}_{\bar{\Lambda}}
= \frac{1}{2}\left(
m_1^2 + m_3^2 + 2 E_1 E_3 - M_{13}^2
\right),
\end{equation}
with the labels $\Lambda(1)$, $\eta(2)$, $\bar \Lambda(3)$ with $M_{ij}$ the invariant mass of particles $i, j$. We also have
\begin{equation}
|\bm{p}_\Lambda| =
\frac{\lambda^{1/2}(m_\psi^2, M_\Lambda^2, M_{23}^2)}{2m_\psi},
\quad
|\bm{p}_{\bar{\Lambda}}| =
\frac{\lambda^{1/2}(m_\psi^2, M^2_{\bar{\Lambda}}, M_{12}^2)}{2m_\psi},
\end{equation}
with
\begin{equation}
 E_1 = \frac{m_{\psi}^2 + M_{\Lambda}^2 - M_{23}^2 }{2 m_\psi},
\quad
 E_3 = \frac{m_{\psi}^2 + M_{\Lambda}^2 - M_{12}^2 }{2 m_\psi},
\end{equation}
and we use the relation
\begin{equation}
M_{12}^2 + M_{13}^2 + M_{23}^2
= m_\psi^2 + M_\Lambda^2 + M^2_{\bar{\Lambda}} + m_\eta^2,
\end{equation}
which allows us to use the PDG mass distribution formula,
in our notation of the fields of Mandl and Shaw~\cite{MandlShaw},
\begin{equation}
\frac{d^2 \Gamma}{dM_{12}\,dM_{23}}
= \frac{1}{(2\pi)^3}
\frac{2 M_{\Lambda} \, 2 M_{\bar{\Lambda}}}{32 m_\psi^3}
\overline{\sum}\sum |t|^2 \, 2M_{12} \, 2M_{23}.
\end{equation}
Integrating over $M_{23}$ within the limits of the PDG, we obtain $d\Gamma/dM_{12}$, and permuting cyclically the indices, we can obtain the three invariant mass distributions $d\Gamma/dM_{ij}$.

The amplitudes $t_{BP, \eta\Lambda}$ that we use are obtained within the chiral unitary approach of Refs.~\cite{Oset:1997it,Oset:2001cn}, where the coupled channels considered are $\bar{K}N$, $\pi\Sigma$, $\eta\Lambda$, $K \Xi$ and the $T$ matrix is given by
\begin{equation}
T = [1 - V G]^{-1} V ,
\end{equation}
with  the interaction kernel $V$ of the type:
\begin{equation}
V_{ij} = -C_{ij}\,\frac{1}{4 f_i f_j}\,(k^0 + k'^0),
\end{equation}
where $k^0$ and $k'^0$ are the energies of the incoming and outgoing mesons in the rest frame of the meson-baryon system. The coefficients $C_{ij}$ are given in Ref.~\cite{Oset:1997it}, and $f_i$ denotes the meson decay constant.
The loop functions are regularized with a cutoff $q_{\rm max}$, which in Ref.~\cite{Oset:1997it} is taken as $q_{\rm max}=630$ MeV.
We take advantage of the high-resolution experimental data of Refs.~\cite{LHCb:2022sck,Belle:2022cbs} to fine-tune the parameters of the theory, $f_i$ and $q_{\text{max},i}$ for the different channels studied the $\Lambda(1670)$ excitation, which has been done in Ref.~\cite{Duan:2024okk}. We use the same parameters to construct our amplitudes, by means of which the pole position of the $\Lambda(1670)$ was found at
$M_{\Lambda(1670)} - i {\Gamma_{\Lambda(1670)}}/{2} =(1669.40 \pm 1.06)-i(21.46 \pm 1.76)$~MeV.

 The data of Ref.~\cite{BESIII:2022cxi} show a peak for the $\Lambda(1670)$ in the $\eta \Lambda$ mass distribution, and a broader peak around 2200 MeV. Part of this latter peak is a replica in the $\eta \Lambda$ mass distribution of the $\bar \Lambda(1670)$ produced in the $\eta \bar \Lambda$  mass distribution, but the data demands a bigger contribution there. In the PDG there are many $\Lambda$ resonances in that energy region, but they have no reported $\eta \Lambda$ decay, or it is very small as in the case of the $\Lambda(2085)$. These resonances have a high spin and we allow for a combined contribution from some of these resonances by adding incoherently to $|t|^2$ of Eq.~\eqref{eq:t_final} the contribution. 
\begin{equation}
\tilde{C}
\left\{
\left|
\frac{M_\Lambda^2}
     {M_{\rm inv}^2(\eta\Lambda)-M_R^2+iM_R\Gamma_R}
\right|^2
+
\left|
\frac{M_\Lambda^2}
     {M_{\rm inv}^2(\eta\bar \Lambda)-M_R^2+iM_R\Gamma_R}
\right|^2
\right\},
\label{eq:extra_res}
\end{equation}
with $M_R = 2200$~MeV and $\Gamma_R = 200$~MeV, and the coefficient $\tilde{C}$ to be 
determined from a fit to the data.
We refrain from fitting the values of the mass and width of $M_R$ and $\Gamma_R$ in Eq.~\eqref{eq:extra_res} to avoid giving the impression that the data shows the need for a specific resonance. 
The contribution is most probably due to several resonances, and the guess of $M_R$ and $\Gamma_R$ is motivated by the missing strength in Fig.~\ref{fig:dGdMinv12_restree} around the 2200~MeV region. Thus, only the strength $\tilde{C}$ is fitted to the data.

\section{Numerical Results}\label{sec:result}

\begin{figure}[!tb]
\begin{center}
\includegraphics[width=0.99\linewidth]{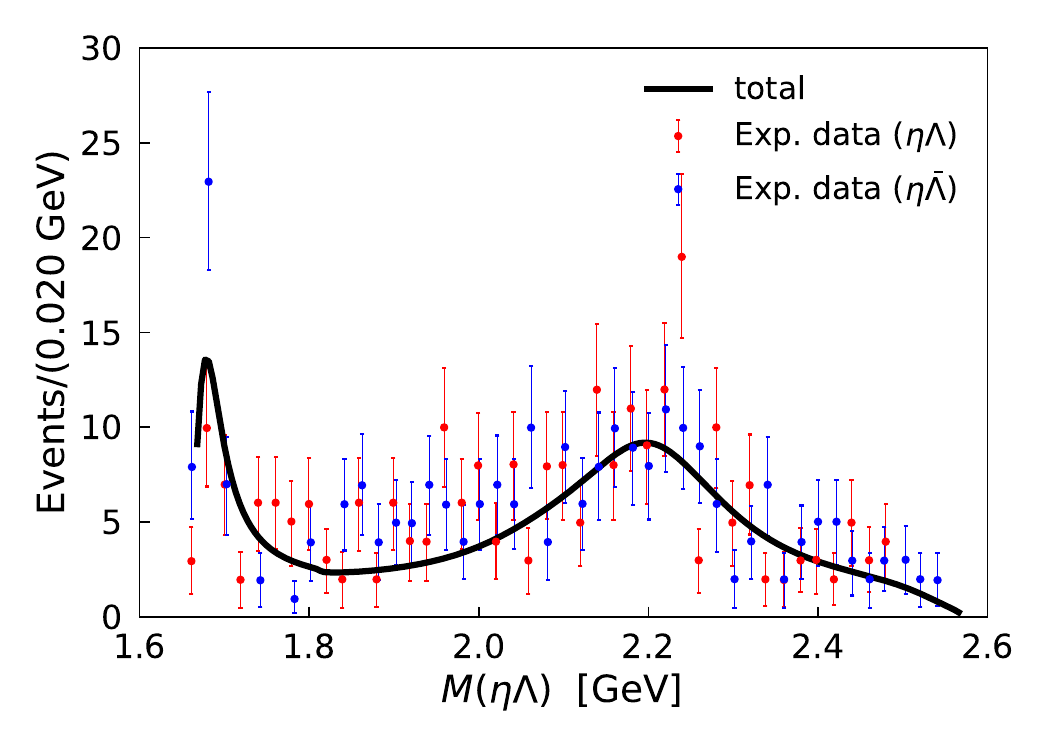}
\caption{ The $\eta \Lambda$ mass distribution in the $\psi(3686) \to \Lambda \bar \Lambda \eta$. The total theoretical result is shown in the black line. 
Experimental data are taken from Ref.~\cite{BESIII:2022cxi}.
}\label{fig:dGdMinv12}
\end{center}
\begin{center}
\includegraphics[width=0.99\linewidth]{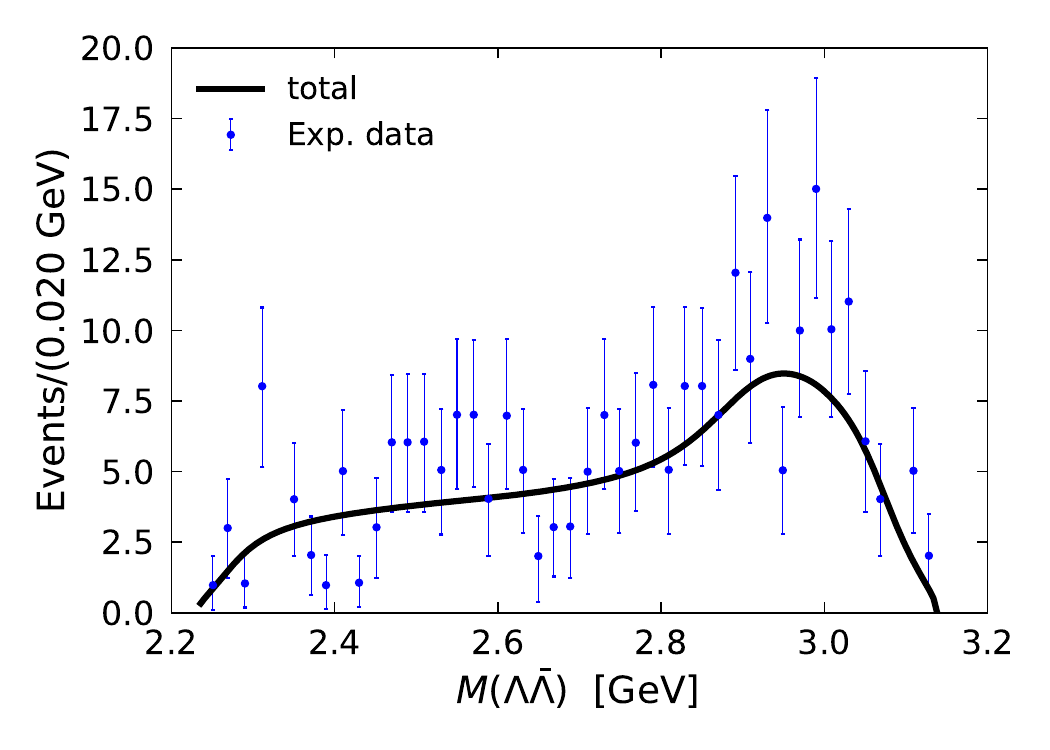}
\caption{
The $\Lambda \bar \Lambda$ mass distribution in the $\psi(3686) \to \Lambda \bar \Lambda \eta$. The total theoretical result is shown in the black line. Experimental data are taken from Ref.~\cite{BESIII:2022cxi}.
}\label{fig:dGdMinv13}
\end{center}
\end{figure}

We show the results for the mass distributions in Figs.~\ref{fig:dGdMinv12}-\ref{fig:dGdMinv13}.
In Fig.~\ref{fig:dGdMinv12}, we present the results for $d\Gamma/dM_{\eta\Lambda}$, which are identical to those for $d\Gamma/dM_{\eta\bar{\Lambda}}$. For this reason, we plot the experimental data for both distributions (in different colors), which are also equal within statistical errors.
We have three parameters, $\tilde{A}$, $\tilde{B}$ and $\tilde{C}$, but one is a global normalization. Hence, we have the ratio $\tilde{A}/\tilde{B}$ and $\tilde{C}$ as free parameters to describe the mass distributions. By adjusting this ratio and $\tilde{C}$, we obtain a fair fit to the data with the values $\tilde{A}/\tilde{B} = 2.80$ and $\tilde{A}^2/\tilde{C} = 241$.
We can see that in the $\eta\Lambda$ mass distribution there are two peaks, a narrow one corresponding to the $\Lambda(1670)$ excitation from $\eta \Lambda$, and a  broader one corresponding to the $\eta \bar{\Lambda}$ excitation when one measures $M_{\eta\Lambda}$.
In addition, there is also the tree-level contribution, the strength of which is tied in our formalism to the $\Lambda(1670)$ excitation.
It is interesting to see the $\Lambda \bar{\Lambda}$ mass distribution, which we show in Fig.~\ref{fig:dGdMinv13}. We observe a broad peak around 3000 MeV, which actually originates from the $\Lambda(1670)$ and $\bar \Lambda(1670)$ excitations appearing in the $\eta \Lambda$ and $\eta \bar{\Lambda}$ mass distributions.
This provides an example of a kinematic reflection of the $\Lambda(1670)$ signal in another invariant-mass projection, sometimes called a replica of the resonance.
The shape of the $\Lambda \bar{\Lambda}$ mass distribution is also fairly well reproduced.

\begin{figure}[!htb]
\begin{center}
\includegraphics[width=0.99\linewidth]{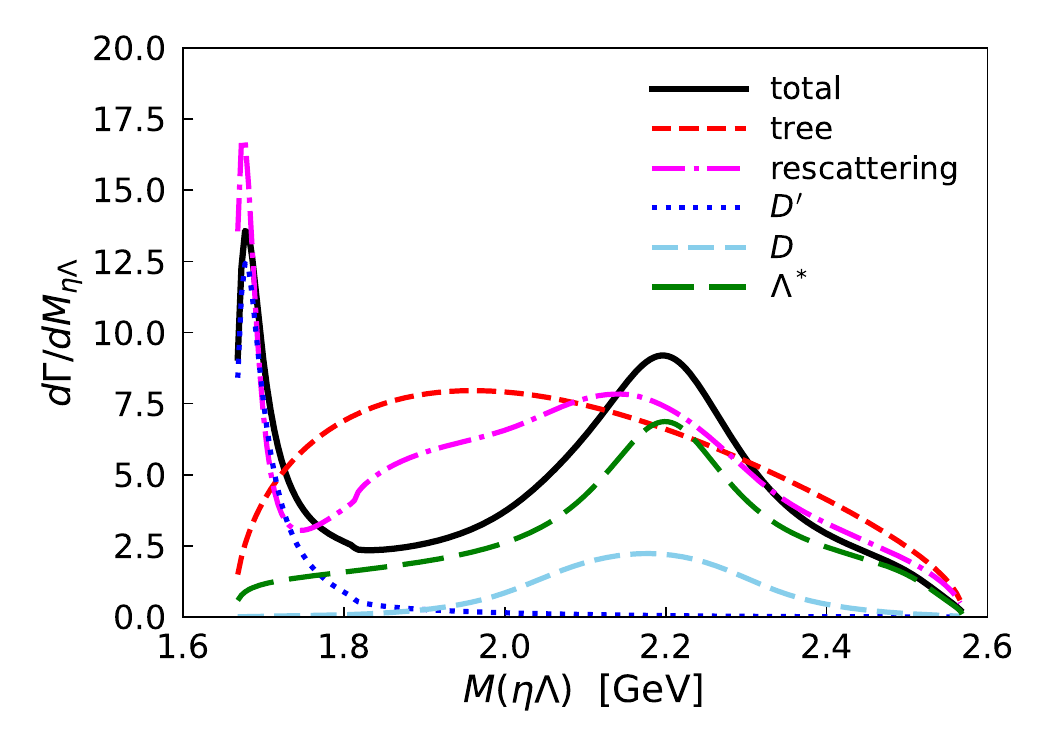}
\caption{
The contributions from the tree-level term, the rescattering term, the cases with only the $D^\prime$ or $D$ terms, and the extra $\Lambda^*$ resonance are shown. The total theoretical result for the $\eta\Lambda$ mass distribution is the same as that shown in Fig.~\ref{fig:dGdMinv12}.
}\label{fig:subcomp12}
\end{center}
\begin{center}
\includegraphics[width=0.99\linewidth]{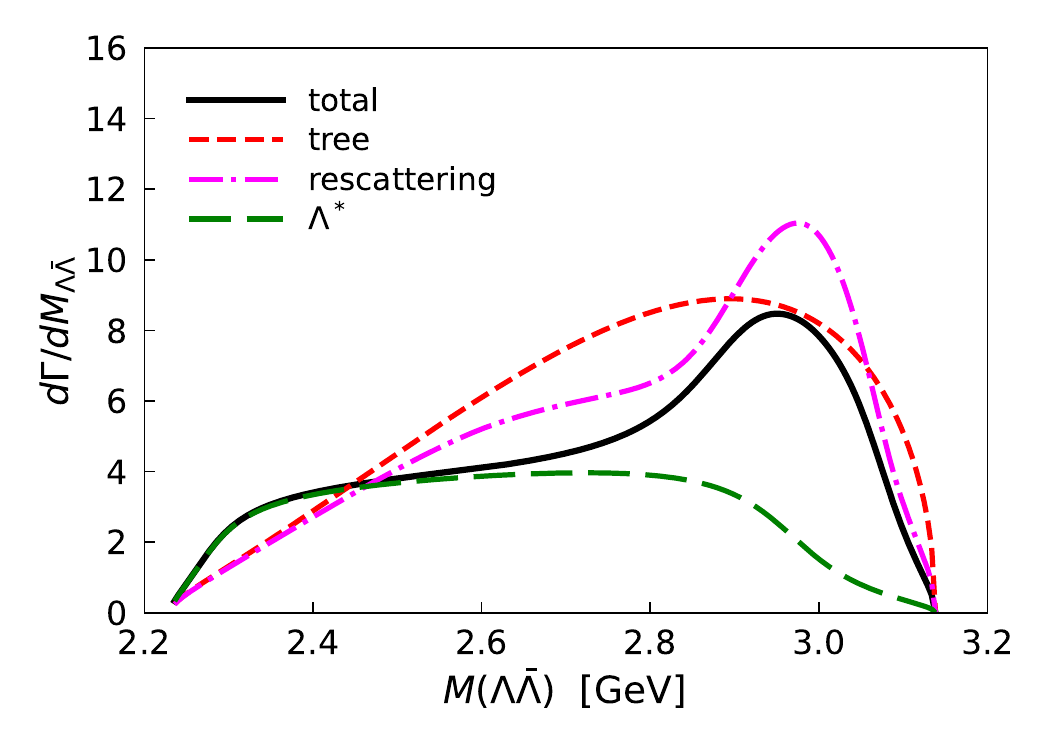}
\caption{
The contributions from the tree-level term, the rescattering term, and the extra $\Lambda^*$ resonance are shown. The total theoretical result for the $\Lambda \bar \Lambda$ mass distribution is the same as that shown in Fig.~\ref{fig:dGdMinv13}.
}\label{fig:subcomp13}
\end{center}
\end{figure}

In order to understand the relevance of the different terms to the process, we show their contributions in Figs.~\ref{fig:subcomp12} and~\ref{fig:subcomp13}. 
The term labelled $D^\prime$ stands for the results where in Eq.~\eqref{eq:D} we take $D=0$ and keep the term $D^\prime$.
 This provides the rescattering terms which generate dynamically the $\Lambda(1670)$, leading to the final $\eta \Lambda$. We can see clearly the shape of the $\Lambda(1670)$ as it appears in the experiment. 
The term labelled $D$ stands for the results where in Eq.~\eqref{eq:Dp} we take $D^\prime =0$ and keep the term $D$.  In this case we obtain a broad contribution in the $\eta \Lambda$ mass distribution around 2200 MeV. 
This is again the kinematic reflection, or replica, of a resonance.  
The resonance with these terms appears for the  $\bar \Lambda(1670)$  and shows in the $\eta \Lambda$ mass distribution in terms of this broad peak. This peak appears in the same region as in the experimental data, as seen in Fig.~\ref{fig:dGdMinv12}, however, it is short of providing the strength shown by the experiment. This is why the data are demanding a contribution from some other sources that we have added phenomenologically by means of Eq.~\eqref{eq:extra_res}.  In the figure we also see that the tree level is not small, and, as seen in Eq.~\eqref{eq:D}, it adds coherently to the rescattering terms.  Finally,  we also show the contribution of the combined extra $\Lambda^*$ resonances of Eq.~\eqref{eq:extra_res}, which, as can be seen, is also not small.  Altogether, the $\eta \Lambda$ mass distribution is well reproduced, as seen in Fig.~\ref{fig:dGdMinv12}, and we show that the contribution of the $\Lambda(1670)$ comes from the rescattering terms, as it should be for a dynamically generated resonance that we have assumed here. 
   The information on Fig.~\ref{fig:subcomp13} on the $\Lambda \bar \Lambda$ mass distribution is also relevant. We separate the tree level contribution to show that there is some structure in the experimental mass distribution. Indeed, there is a peak around 3000 MeV which comes from the $\Lambda(1670)$ and  $\bar \Lambda(1670)$ contributions, but again, we can see that this contribution alone would not provide the line shape of the mass distributions of Fig.~\ref{fig:dGdMinv13}. What we see there is that the contribution of the extra $\Lambda^*$ resonances, via Eq.~\eqref{eq:extra_res}, fills up the region of low invariant masses to give a good reproduction of the data of Fig.~\ref{fig:dGdMinv13}. 
   
\begin{figure}[!htb]
\begin{center}
\includegraphics[width=0.99\linewidth]{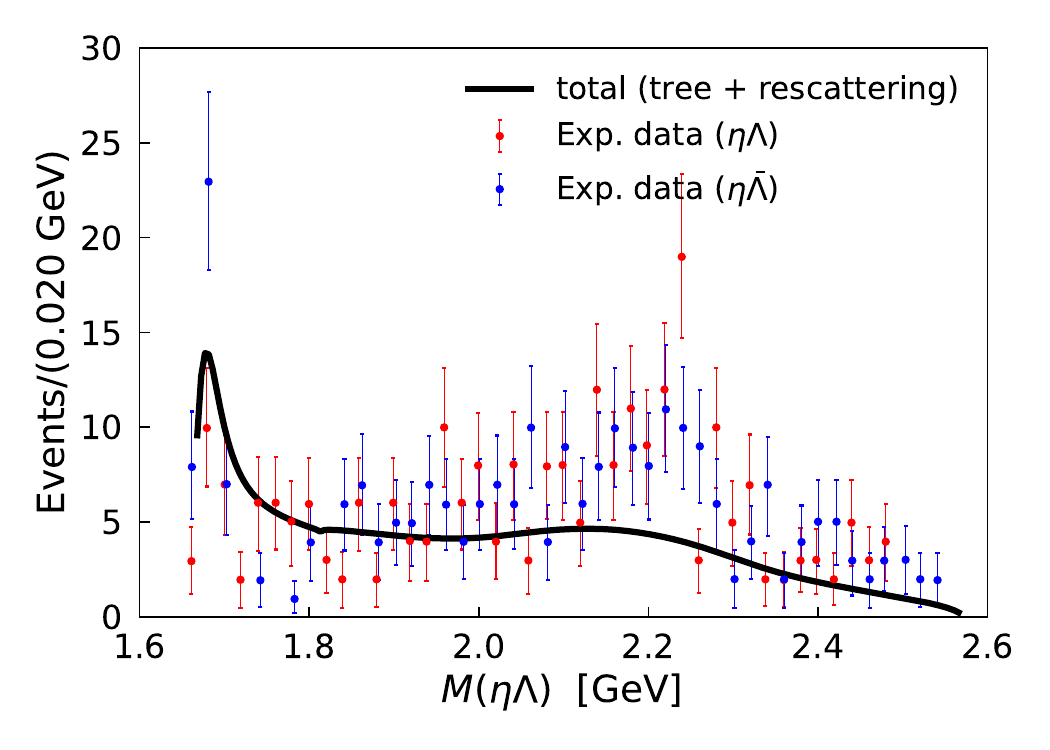}
\caption{
The same as Fig.~\ref{fig:dGdMinv12}, without the extra $\Lambda^*$ resonance term.
}\label{fig:dGdMinv12_restree}
\end{center}
\begin{center}
\includegraphics[width=0.99\linewidth]{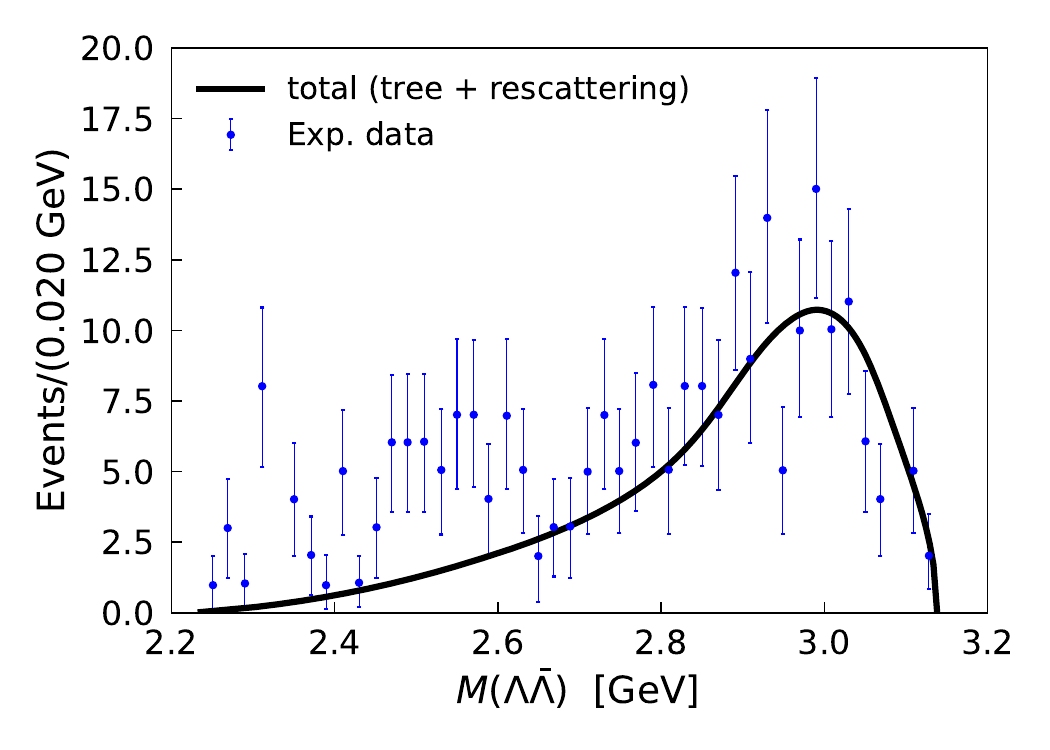}
\caption{ 
The same as Fig.~\ref{fig:dGdMinv13}, without the extra $\Lambda^*$ resonance term.
}\label{fig:dGdMinv13_restree}
\end{center}
\end{figure}

   To further support our claims of a needed extra $\Lambda^*$ contribution, we show in Fig.~\ref{fig:dGdMinv12_restree} the results for the $\eta \Lambda$ mass distribution without the $\Lambda^*$ term of Eq.~\eqref{eq:extra_res}. The best fit to the data shows clearly that there is some missing strength in the second peak. Similarly, in Fig.~\ref{fig:dGdMinv13_restree} we show the results of the former fit for the data in the  $\Lambda \bar \Lambda$ mass distribution and we see that there is missing strength in the lower region of the mass spectrum.   
We can make the differences between both fits more quantitative. The fit without the extra $\Lambda^*$ contribution of Figs.~\ref{fig:dGdMinv12_restree} and \ref{fig:dGdMinv13_restree} has a reduced chi-square, $\chi^2/{\rm dof}=1.87$, 
while the fit shown in Figs.~\ref{fig:dGdMinv12} and \ref{fig:dGdMinv13}, including the extra $\Lambda^*$  contribution, has $\chi^2/{\rm dof}=1.22$.
The parameters $\tilde{A}$ and $\tilde{B}$ are strongly correlated, with a correlation coefficient of $\rho_{12} \simeq 0.96$.
Therefore, we do not attach much attention to these individual parameters, which are tied to the normalization of the data, but the ratio  $\tilde{A}/\tilde{B}$ is more stable and we obtain
$\tilde{A}/\tilde{B} = 2.0 \pm 0.1$ for the fit of Figs.~\ref{fig:dGdMinv12_restree} and \ref{fig:dGdMinv13_restree}, and
$\tilde{A}/\tilde{B} = 2.8 \pm 0.2$ for the fit of Figs.~\ref{fig:dGdMinv12} and \ref{fig:dGdMinv13}.
The value of $\tilde{C}$ in the fit to Figs.~\ref{fig:dGdMinv12} and \ref{fig:dGdMinv13} is $\tilde{C}=(3.3\pm0.2)\times10^{-3}$.
We can see that the value of $\tilde{A}/\tilde{B}$ is stable within statistical uncertainties, and the fit improves appreciably when accounting for possible extra $\Lambda^*$ contributions.
These observations should provide further motivation for continued experimental studies of the decay channels of the $\Lambda^*$ resonances, particularly the $\eta \Lambda$ decay channel.

At this point, we would like to make some comments on the work of Ref.~\cite{He:2026mkf}, where they use the flavor structure of Eq.~\eqref{eq:Lagrangian}. The $\tilde{D}$ and $\tilde{F}$ parameters are related to our $\tilde{A}$ and $\tilde{B}$. Concretely,
\begin{equation}
\tilde{A} = \tilde{D} + \tilde{F}; \qquad
\tilde{B} = \tilde{D} - \tilde{F}.
\end{equation}
With our result $\tilde{A} = 2.8\tilde{B}$, we obtain ${\tilde{F}}/{\tilde{D}} = 0.47$.
This should be compared to the value $0.50 \pm 0.06$ reported in Ref.~\cite{He:2026mkf} from the study of $\psi(3686)$ decays into $\Lambda \bar{\Sigma} \pi$, $\Lambda \bar{N} \pi$, $\Lambda \bar{\Lambda} \eta$, and $\Sigma \bar{N} \pi$ at the tree level.
Given that, as we have seen, the resonant contribution is of the same order of magnitude as the tree-level contribution, and that similar effects can be expected in the other reactions mentioned above, the value of $\tilde{F}/\tilde{D}$ extracted in Ref.~\cite{He:2026mkf} should be regarded as only indicative rather than precise~\cite{Guo_private}. Nevertheless, the agreement obtained is most encouraging.

\section {Conclusions}\label{sec:discussion}
We have made a theoretical study of the $\psi(3686) \to  \Lambda \bar \Lambda \eta$ reaction, making use of the fact that the $\psi(3686)$ state, corresponding to a $c \bar c$ state, is a singlet of SU(3) in the $u,d,s$ quarks sector. This provides two flavor structures corresponding to two different traces that one can construct with the product of the meson ($P$), baryon ($B$) and antibaryon ($\bar B)$ SU(3) matrices. Up to a global normalization, the relative weight of these two structures is the only free parameter that we have to reproduce the experimental numbers. We show that the two former structures give rise to $\Lambda \bar \Lambda \eta$ at the tree level, but there are other $\bar B$, $B$, $P$ combinations that also lead to $\Lambda \bar \Lambda \eta$ after rescattering of a meson and a baryon. These transition matrices contain the information about the $\Lambda(1670)$, which in our approach is dynamically generated by the interaction of pseudoscalar mesons with baryons. We obtain mass distributions in $\eta \Lambda$ and $\eta \bar \Lambda$ with a distinct peak in the region of the  $\Lambda(1670)$ and $\bar \Lambda(1670)$, respectively.

   Up to a global normalization adjusted to the number of events in the experiment, our results depend on the $\tilde A / \tilde B $ ratio and the $\tilde C$ parameter, and we find a good description of the two mass distributions with a value of the $\tilde A / \tilde B $ ratio consistent with values obtained from a global analysis of several reactions. We observed that the tree level contribution and that of the $\Lambda(1670)$ resonance are of the same order of magnitude, and there is some interference between them. We also observed that with the tree level and the resonance contributions alone we could not get a quantitative description of the data, which demanded extra resonant strength around 2200 MeV in the $\eta \Lambda$ mass distribution.

Our results, together with those obtained for the production of the $\Lambda(1670)$ in several other reactions discussed in the Introduction, provide further support for the dynamical origin of the $\Lambda(1670)$ as a state generated by coupled-channel interactions, as well as for its close connection to the higher-energy pole of the $\Lambda(1405)$, commonly denoted as $\Lambda(1420)$.

   The other relevant finding of the work was the need for extra resonant contribution 
 in the region of mass of 2200 MeV.  There are several resonances tabulated in the PDG around this mass, but the  $\eta \Lambda$ decay channel is not observed, or, in only one case where it has been observed, it is small. The findings in this work should motivate further search of this decay channel of those resonances.

\section*{Acknowledgments}
The work was partly supported by JSPS KAKENHI Grant Number JP24K07020.
This work is partly supported by the Spanish Ministerio de Economia y Competitividad (MINECO) and European FEDER funds under Contracts No. FIS2017-84038-C2-1-P B, PID2020- 112777GB-I00, and by Generalitat Valenciana under contracts PROMETEO/2020/023 and CIPROM/2023/59.This project has received funding from the European Union Horizon 2020 research and innovation programme under the program H2020- INFRAIA2018-1, grant agreement No. 824093 of the STRONG-2020 project. This work is supported by the Spanish Ministerio de Ciencia e Innovacion (MICINN) under contracts PID2020-112777GB-I00, PID2023-147458NBC21 and CEX2023-001292-S.

\bibliography{ref_Lambda.bib}

\begin{thebibliography}{30}%
\makeatletter
\providecommand \@ifxundefined [1]{%
 \@ifx{#1\undefined}
}%
\providecommand \@ifnum [1]{%
 \ifnum #1\expandafter \@firstoftwo
 \else \expandafter \@secondoftwo
 \fi
}%
\providecommand \@ifx [1]{%
 \ifx #1\expandafter \@firstoftwo
 \else \expandafter \@secondoftwo
 \fi
}%
\providecommand \natexlab [1]{#1}%
\providecommand \enquote  [1]{``#1''}%
\providecommand \bibnamefont  [1]{#1}%
\providecommand \bibfnamefont [1]{#1}%
\providecommand \citenamefont [1]{#1}%
\providecommand \href@noop [0]{\@secondoftwo}%
\providecommand \href [0]{\begingroup \@sanitize@url \@href}%
\providecommand \@href[1]{\@@startlink{#1}\@@href}%
\providecommand \@@href[1]{\endgroup#1\@@endlink}%
\providecommand \@sanitize@url [0]{\catcode `\\12\catcode `\$12\catcode
  `\&12\catcode `\#12\catcode `\^12\catcode `\_12\catcode `\%12\relax}%
\providecommand \@@startlink[1]{}%
\providecommand \@@endlink[0]{}%
\providecommand \url  [0]{\begingroup\@sanitize@url \@url }%
\providecommand \@url [1]{\endgroup\@href {#1}{\urlprefix }}%
\providecommand \urlprefix  [0]{URL }%
\providecommand \Eprint [0]{\href }%
\providecommand \doibase [0]{http://dx.doi.org/}%
\providecommand \selectlanguage [0]{\@gobble}%
\providecommand \bibinfo  [0]{\@secondoftwo}%
\providecommand \bibfield  [0]{\@secondoftwo}%
\providecommand \translation [1]{[#1]}%
\providecommand \BibitemOpen [0]{}%
\providecommand \bibitemStop [0]{}%
\providecommand \bibitemNoStop [0]{.\EOS\space}%
\providecommand \EOS [0]{\spacefactor3000\relax}%
\providecommand \BibitemShut  [1]{\csname bibitem#1\endcsname}%
\let\auto@bib@innerbib\@empty
\bibitem [{\citenamefont {Gopal}\ \emph {et~al.}(1977)\citenamefont {Gopal},
  \citenamefont {Ross}, \citenamefont {Van~Horn}, \citenamefont {McPherson},
  \citenamefont {Clayton}, \citenamefont {Bacon},\ and\ \citenamefont
  {Butterworth}}]{Gopal:1976gs}%
  \BibitemOpen
  \bibfield  {author} {\bibinfo {author} {\bibfnamefont {G.~P.}\ \bibnamefont
  {Gopal}}, \bibinfo {author} {\bibfnamefont {R.~T.}\ \bibnamefont {Ross}},
  \bibinfo {author} {\bibfnamefont {A.~J.}\ \bibnamefont {Van~Horn}}, \bibinfo
  {author} {\bibfnamefont {A.~C.}\ \bibnamefont {McPherson}}, \bibinfo {author}
  {\bibfnamefont {E.~F.}\ \bibnamefont {Clayton}}, \bibinfo {author}
  {\bibfnamefont {T.~C.}\ \bibnamefont {Bacon}}, \ and\ \bibinfo {author}
  {\bibfnamefont {I.}~\bibnamefont {Butterworth}} (\bibinfo {collaboration}
  {Rutherford-London}),\ }\href {\doibase 10.1016/0550-3213(77)90002-5}
  {\bibfield  {journal} {\bibinfo  {journal} {Nucl. Phys. B}\ }\textbf
  {\bibinfo {volume} {119}},\ \bibinfo {pages} {362} (\bibinfo {year}
  {1977})}\BibitemShut {NoStop}%
\bibitem [{\citenamefont {Lee}\ \emph {et~al.}(2021)\citenamefont {Lee} \emph
  {et~al.}}]{Belle:2020xku}%
  \BibitemOpen
  \bibfield  {author} {\bibinfo {author} {\bibfnamefont {J.~Y.}\ \bibnamefont
  {Lee}} \emph {et~al.} (\bibinfo {collaboration} {Belle}),\ }\href {\doibase
  10.1103/PhysRevD.103.052005} {\bibfield  {journal} {\bibinfo  {journal}
  {Phys. Rev. D}\ }\textbf {\bibinfo {volume} {103}},\ \bibinfo {pages}
  {052005} (\bibinfo {year} {2021})},\ \Eprint
  {http://arxiv.org/abs/2008.11575} {arXiv:2008.11575 [hep-ex]} \BibitemShut
  {NoStop}%
\bibitem [{\citenamefont {Ablikim}\ \emph {et~al.}(2026)\citenamefont {Ablikim}
  \emph {et~al.}}]{BESIII:2026vrm}%
  \BibitemOpen
  \bibfield  {author} {\bibinfo {author} {\bibfnamefont {M.}~\bibnamefont
  {Ablikim}} \emph {et~al.} (\bibinfo {collaboration} {BESIII}),\ }\href@noop
  {} {\  (\bibinfo {year} {2026})},\ \Eprint {http://arxiv.org/abs/2601.07617}
  {arXiv:2601.07617 [hep-ex]} \BibitemShut {NoStop}%
\bibitem [{\citenamefont {Yang}\ \emph {et~al.}(2023)\citenamefont {Yang} \emph
  {et~al.}}]{Belle:2022cbs}%
  \BibitemOpen
  \bibfield  {author} {\bibinfo {author} {\bibfnamefont {S.~B.}\ \bibnamefont
  {Yang}} \emph {et~al.} (\bibinfo {collaboration} {Belle}),\ }\href {\doibase
  10.1103/PhysRevD.108.L031104} {\bibfield  {journal} {\bibinfo  {journal}
  {Phys. Rev. D}\ }\textbf {\bibinfo {volume} {108}},\ \bibinfo {pages}
  {L031104} (\bibinfo {year} {2023})},\ \Eprint
  {http://arxiv.org/abs/2209.00050} {arXiv:2209.00050 [hep-ex]} \BibitemShut
  {NoStop}%
\bibitem [{\citenamefont {Aaij}\ \emph {et~al.}(2023)\citenamefont {Aaij} \emph
  {et~al.}}]{LHCb:2022sck}%
  \BibitemOpen
  \bibfield  {author} {\bibinfo {author} {\bibfnamefont {R.}~\bibnamefont
  {Aaij}} \emph {et~al.} (\bibinfo {collaboration} {LHCb}),\ }\href {\doibase
  10.1103/PhysRevD.108.012023} {\bibfield  {journal} {\bibinfo  {journal}
  {Phys. Rev. D}\ }\textbf {\bibinfo {volume} {108}},\ \bibinfo {pages}
  {012023} (\bibinfo {year} {2023})},\ \Eprint
  {http://arxiv.org/abs/2208.03262} {arXiv:2208.03262 [hep-ex]} \BibitemShut
  {NoStop}%
\bibitem [{\citenamefont {Oset}\ \emph {et~al.}(2002)\citenamefont {Oset},
  \citenamefont {Ramos},\ and\ \citenamefont {Bennhold}}]{Oset:2001cn}%
  \BibitemOpen
  \bibfield  {author} {\bibinfo {author} {\bibfnamefont {E.}~\bibnamefont
  {Oset}}, \bibinfo {author} {\bibfnamefont {A.}~\bibnamefont {Ramos}}, \ and\
  \bibinfo {author} {\bibfnamefont {C.}~\bibnamefont {Bennhold}},\ }\href
  {\doibase 10.1016/S0370-2693(01)01523-4} {\bibfield  {journal} {\bibinfo
  {journal} {Phys. Lett. B}\ }\textbf {\bibinfo {volume} {527}},\ \bibinfo
  {pages} {99} (\bibinfo {year} {2002})},\ \bibinfo {note} {[Erratum:
  Phys.Lett.B 530, 260--260 (2002)]},\ \Eprint
  {http://arxiv.org/abs/nucl-th/0109006} {arXiv:nucl-th/0109006} \BibitemShut
  {NoStop}%
\bibitem [{\citenamefont {Garcia-Recio}\ \emph {et~al.}(2003)\citenamefont
  {Garcia-Recio}, \citenamefont {Nieves}, \citenamefont {Ruiz~Arriola},\ and\
  \citenamefont {Vicente~Vacas}}]{Garcia-Recio:2002yxy}%
  \BibitemOpen
  \bibfield  {author} {\bibinfo {author} {\bibfnamefont {C.}~\bibnamefont
  {Garcia-Recio}}, \bibinfo {author} {\bibfnamefont {J.}~\bibnamefont
  {Nieves}}, \bibinfo {author} {\bibfnamefont {E.}~\bibnamefont
  {Ruiz~Arriola}}, \ and\ \bibinfo {author} {\bibfnamefont {M.~J.}\
  \bibnamefont {Vicente~Vacas}},\ }\href {\doibase 10.1103/PhysRevD.67.076009}
  {\bibfield  {journal} {\bibinfo  {journal} {Phys. Rev. D}\ }\textbf {\bibinfo
  {volume} {67}},\ \bibinfo {pages} {076009} (\bibinfo {year} {2003})},\
  \Eprint {http://arxiv.org/abs/hep-ph/0210311} {arXiv:hep-ph/0210311}
  \BibitemShut {NoStop}%
\bibitem [{\citenamefont {Kolomeitsev}\ and\ \citenamefont
  {Lutz}(2004)}]{Kolomeitsev:2003kt}%
  \BibitemOpen
  \bibfield  {author} {\bibinfo {author} {\bibfnamefont {E.~E.}\ \bibnamefont
  {Kolomeitsev}}\ and\ \bibinfo {author} {\bibfnamefont {M.~F.~M.}\
  \bibnamefont {Lutz}},\ }\href {\doibase 10.1016/j.physletb.2004.01.066}
  {\bibfield  {journal} {\bibinfo  {journal} {Phys. Lett. B}\ }\textbf
  {\bibinfo {volume} {585}},\ \bibinfo {pages} {243} (\bibinfo {year}
  {2004})},\ \Eprint {http://arxiv.org/abs/nucl-th/0305101}
  {arXiv:nucl-th/0305101} \BibitemShut {NoStop}%
\bibitem [{\citenamefont {Oller}(2006)}]{Oller:2006jw}%
  \BibitemOpen
  \bibfield  {author} {\bibinfo {author} {\bibfnamefont {J.~A.}\ \bibnamefont
  {Oller}},\ }\href {\doibase 10.1140/epja/i2006-10011-3} {\bibfield  {journal}
  {\bibinfo  {journal} {Eur. Phys. J. A}\ }\textbf {\bibinfo {volume} {28}},\
  \bibinfo {pages} {63} (\bibinfo {year} {2006})},\ \Eprint
  {http://arxiv.org/abs/hep-ph/0603134} {arXiv:hep-ph/0603134} \BibitemShut
  {NoStop}%
\bibitem [{\citenamefont {Oset}\ and\ \citenamefont
  {Ramos}(1998)}]{Oset:1997it}%
  \BibitemOpen
  \bibfield  {author} {\bibinfo {author} {\bibfnamefont {E.}~\bibnamefont
  {Oset}}\ and\ \bibinfo {author} {\bibfnamefont {A.}~\bibnamefont {Ramos}},\
  }\href {\doibase 10.1016/S0375-9474(98)00170-5} {\bibfield  {journal}
  {\bibinfo  {journal} {Nucl. Phys. A}\ }\textbf {\bibinfo {volume} {635}},\
  \bibinfo {pages} {99} (\bibinfo {year} {1998})},\ \Eprint
  {http://arxiv.org/abs/nucl-th/9711022} {arXiv:nucl-th/9711022} \BibitemShut
  {NoStop}%
\bibitem [{\citenamefont {Jido}\ \emph {et~al.}(2003)\citenamefont {Jido},
  \citenamefont {Oller}, \citenamefont {Oset}, \citenamefont {Ramos},\ and\
  \citenamefont {Meissner}}]{Jido:2003cb}%
  \BibitemOpen
  \bibfield  {author} {\bibinfo {author} {\bibfnamefont {D.}~\bibnamefont
  {Jido}}, \bibinfo {author} {\bibfnamefont {J.~A.}\ \bibnamefont {Oller}},
  \bibinfo {author} {\bibfnamefont {E.}~\bibnamefont {Oset}}, \bibinfo {author}
  {\bibfnamefont {A.}~\bibnamefont {Ramos}}, \ and\ \bibinfo {author}
  {\bibfnamefont {U.~G.}\ \bibnamefont {Meissner}},\ }\href {\doibase
  10.1016/S0375-9474(03)01598-7} {\bibfield  {journal} {\bibinfo  {journal}
  {Nucl. Phys. A}\ }\textbf {\bibinfo {volume} {725}},\ \bibinfo {pages} {181}
  (\bibinfo {year} {2003})},\ \Eprint {http://arxiv.org/abs/nucl-th/0303062}
  {arXiv:nucl-th/0303062} \BibitemShut {NoStop}%
\bibitem [{\citenamefont {Navas}\ \emph {et~al.}(2024)\citenamefont {Navas}
  \emph {et~al.}}]{ParticleDataGroup:2024cfk}%
  \BibitemOpen
  \bibfield  {author} {\bibinfo {author} {\bibfnamefont {S.}~\bibnamefont
  {Navas}} \emph {et~al.} (\bibinfo {collaboration} {Particle Data Group}),\
  }\href {\doibase 10.1103/PhysRevD.110.030001} {\bibfield  {journal} {\bibinfo
   {journal} {Phys. Rev. D}\ }\textbf {\bibinfo {volume} {110}},\ \bibinfo
  {pages} {030001} (\bibinfo {year} {2024})}\BibitemShut {NoStop}%
\bibitem [{\citenamefont {Starostin}\ \emph {et~al.}(2001)\citenamefont
  {Starostin} \emph {et~al.}}]{CrystalBall:2001uhc}%
  \BibitemOpen
  \bibfield  {author} {\bibinfo {author} {\bibfnamefont {A.}~\bibnamefont
  {Starostin}} \emph {et~al.} (\bibinfo {collaboration} {Crystal Ball}),\
  }\href {\doibase 10.1103/PhysRevC.64.055205} {\bibfield  {journal} {\bibinfo
  {journal} {Phys. Rev. C}\ }\textbf {\bibinfo {volume} {64}},\ \bibinfo
  {pages} {055205} (\bibinfo {year} {2001})}\BibitemShut {NoStop}%
\bibitem [{\citenamefont {Zhong}\ and\ \citenamefont
  {Zhao}(2009)}]{Zhong:2008km}%
  \BibitemOpen
  \bibfield  {author} {\bibinfo {author} {\bibfnamefont {X.-H.}\ \bibnamefont
  {Zhong}}\ and\ \bibinfo {author} {\bibfnamefont {Q.}~\bibnamefont {Zhao}},\
  }\href {\doibase 10.1103/PhysRevC.79.045202} {\bibfield  {journal} {\bibinfo
  {journal} {Phys. Rev. C}\ }\textbf {\bibinfo {volume} {79}},\ \bibinfo
  {pages} {045202} (\bibinfo {year} {2009})},\ \Eprint
  {http://arxiv.org/abs/0811.4212} {arXiv:0811.4212 [nucl-th]} \BibitemShut
  {NoStop}%
\bibitem [{\citenamefont {Miyahara}\ \emph {et~al.}(2015)\citenamefont
  {Miyahara}, \citenamefont {Hyodo},\ and\ \citenamefont
  {Oset}}]{Miyahara:2015cja}%
  \BibitemOpen
  \bibfield  {author} {\bibinfo {author} {\bibfnamefont {K.}~\bibnamefont
  {Miyahara}}, \bibinfo {author} {\bibfnamefont {T.}~\bibnamefont {Hyodo}}, \
  and\ \bibinfo {author} {\bibfnamefont {E.}~\bibnamefont {Oset}},\ }\href
  {\doibase 10.1103/PhysRevC.92.055204} {\bibfield  {journal} {\bibinfo
  {journal} {Phys. Rev. C}\ }\textbf {\bibinfo {volume} {92}},\ \bibinfo
  {pages} {055204} (\bibinfo {year} {2015})},\ \Eprint
  {http://arxiv.org/abs/1508.04882} {arXiv:1508.04882 [nucl-th]} \BibitemShut
  {NoStop}%
\bibitem [{\citenamefont {Xie}\ and\ \citenamefont {Geng}(2016)}]{Xie:2016evi}%
  \BibitemOpen
  \bibfield  {author} {\bibinfo {author} {\bibfnamefont {J.-J.}\ \bibnamefont
  {Xie}}\ and\ \bibinfo {author} {\bibfnamefont {L.-S.}\ \bibnamefont {Geng}},\
  }\href {\doibase 10.1140/epjc/s10052-016-4342-z} {\bibfield  {journal}
  {\bibinfo  {journal} {Eur. Phys. J. C}\ }\textbf {\bibinfo {volume} {76}},\
  \bibinfo {pages} {496} (\bibinfo {year} {2016})},\ \Eprint
  {http://arxiv.org/abs/1604.02756} {arXiv:1604.02756 [nucl-th]} \BibitemShut
  {NoStop}%
\bibitem [{\citenamefont {Wang}\ \emph {et~al.}(2022)\citenamefont {Wang},
  \citenamefont {Wei}, \citenamefont {Yang}, \citenamefont {Wang},
  \citenamefont {Geng},\ and\ \citenamefont {Xie}}]{Wang:2022nac}%
  \BibitemOpen
  \bibfield  {author} {\bibinfo {author} {\bibfnamefont {G.-Y.}\ \bibnamefont
  {Wang}}, \bibinfo {author} {\bibfnamefont {N.-C.}\ \bibnamefont {Wei}},
  \bibinfo {author} {\bibfnamefont {H.-M.}\ \bibnamefont {Yang}}, \bibinfo
  {author} {\bibfnamefont {E.}~\bibnamefont {Wang}}, \bibinfo {author}
  {\bibfnamefont {L.-S.}\ \bibnamefont {Geng}}, \ and\ \bibinfo {author}
  {\bibfnamefont {J.-J.}\ \bibnamefont {Xie}},\ }\href {\doibase
  10.1103/PhysRevD.106.056001} {\bibfield  {journal} {\bibinfo  {journal}
  {Phys. Rev. D}\ }\textbf {\bibinfo {volume} {106}},\ \bibinfo {pages}
  {056001} (\bibinfo {year} {2022})},\ \Eprint
  {http://arxiv.org/abs/2206.01425} {arXiv:2206.01425 [hep-ph]} \BibitemShut
  {NoStop}%
\bibitem [{\citenamefont {Duan}\ \emph {et~al.}(2025)\citenamefont {Duan},
  \citenamefont {Lyu}, \citenamefont {Xiao}, \citenamefont {Wang},
  \citenamefont {Xie}, \citenamefont {Chen},\ and\ \citenamefont
  {Oset}}]{Duan:2024czu}%
  \BibitemOpen
  \bibfield  {author} {\bibinfo {author} {\bibfnamefont {M.-Y.}\ \bibnamefont
  {Duan}}, \bibinfo {author} {\bibfnamefont {W.-T.}\ \bibnamefont {Lyu}},
  \bibinfo {author} {\bibfnamefont {C.-W.}\ \bibnamefont {Xiao}}, \bibinfo
  {author} {\bibfnamefont {E.}~\bibnamefont {Wang}}, \bibinfo {author}
  {\bibfnamefont {J.-J.}\ \bibnamefont {Xie}}, \bibinfo {author} {\bibfnamefont
  {D.-Y.}\ \bibnamefont {Chen}}, \ and\ \bibinfo {author} {\bibfnamefont
  {E.}~\bibnamefont {Oset}},\ }\href {\doibase 10.1103/PhysRevD.111.016004}
  {\bibfield  {journal} {\bibinfo  {journal} {Phys. Rev. D}\ }\textbf {\bibinfo
  {volume} {111}},\ \bibinfo {pages} {016004} (\bibinfo {year} {2025})},\
  \Eprint {http://arxiv.org/abs/2410.16078} {arXiv:2410.16078 [hep-ph]}
  \BibitemShut {NoStop}%
\bibitem [{\citenamefont {Lyu}\ \emph {et~al.}(2024)\citenamefont {Lyu},
  \citenamefont {Zhang}, \citenamefont {Wang}, \citenamefont {Wu},
  \citenamefont {Wang}, \citenamefont {Geng},\ and\ \citenamefont
  {Xie}}]{Lyu:2024qgc}%
  \BibitemOpen
  \bibfield  {author} {\bibinfo {author} {\bibfnamefont {W.-T.}\ \bibnamefont
  {Lyu}}, \bibinfo {author} {\bibfnamefont {S.-C.}\ \bibnamefont {Zhang}},
  \bibinfo {author} {\bibfnamefont {G.-Y.}\ \bibnamefont {Wang}}, \bibinfo
  {author} {\bibfnamefont {J.-J.}\ \bibnamefont {Wu}}, \bibinfo {author}
  {\bibfnamefont {E.}~\bibnamefont {Wang}}, \bibinfo {author} {\bibfnamefont
  {L.-S.}\ \bibnamefont {Geng}}, \ and\ \bibinfo {author} {\bibfnamefont
  {J.-J.}\ \bibnamefont {Xie}},\ }\href {\doibase 10.1103/PhysRevD.110.054020}
  {\bibfield  {journal} {\bibinfo  {journal} {Phys. Rev. D}\ }\textbf {\bibinfo
  {volume} {110}},\ \bibinfo {pages} {054020} (\bibinfo {year} {2024})},\
  \Eprint {http://arxiv.org/abs/2405.09226} {arXiv:2405.09226 [hep-ph]}
  \BibitemShut {NoStop}%
\bibitem [{\citenamefont {Lyu}\ \emph {et~al.}(2026)\citenamefont {Lyu},
  \citenamefont {Liu}, \citenamefont {Wu}, \citenamefont {Li},\ and\
  \citenamefont {Wang}}]{Lyu:2026ack}%
  \BibitemOpen
  \bibfield  {author} {\bibinfo {author} {\bibfnamefont {W.-T.}\ \bibnamefont
  {Lyu}}, \bibinfo {author} {\bibfnamefont {S.-W.}\ \bibnamefont {Liu}},
  \bibinfo {author} {\bibfnamefont {J.-J.}\ \bibnamefont {Wu}}, \bibinfo
  {author} {\bibfnamefont {D.-M.}\ \bibnamefont {Li}}, \ and\ \bibinfo {author}
  {\bibfnamefont {E.}~\bibnamefont {Wang}},\ }\href@noop {} {\  (\bibinfo
  {year} {2026})},\ \Eprint {http://arxiv.org/abs/2606.04690} {arXiv:2606.04690
  [hep-ph]} \BibitemShut {NoStop}%
\bibitem [{\citenamefont {Duan}\ \emph {et~al.}(2024)\citenamefont {Duan},
  \citenamefont {Bayar},\ and\ \citenamefont {Oset}}]{Duan:2024okk}%
  \BibitemOpen
  \bibfield  {author} {\bibinfo {author} {\bibfnamefont {M.-Y.}\ \bibnamefont
  {Duan}}, \bibinfo {author} {\bibfnamefont {M.}~\bibnamefont {Bayar}}, \ and\
  \bibinfo {author} {\bibfnamefont {E.}~\bibnamefont {Oset}},\ }\href {\doibase
  10.1016/j.physletb.2024.139003} {\bibfield  {journal} {\bibinfo  {journal}
  {Phys. Lett. B}\ }\textbf {\bibinfo {volume} {857}},\ \bibinfo {pages}
  {139003} (\bibinfo {year} {2024})},\ \Eprint
  {http://arxiv.org/abs/2407.01410} {arXiv:2407.01410 [hep-ph]} \BibitemShut
  {NoStop}%
\bibitem [{\citenamefont {Zhang}\ \emph {et~al.}(2024)\citenamefont {Zhang},
  \citenamefont {Duan}, \citenamefont {Lyu}, \citenamefont {Wang},
  \citenamefont {Zhu},\ and\ \citenamefont {Wang}}]{Zhang:2024jby}%
  \BibitemOpen
  \bibfield  {author} {\bibinfo {author} {\bibfnamefont {S.-C.}\ \bibnamefont
  {Zhang}}, \bibinfo {author} {\bibfnamefont {M.-Y.}\ \bibnamefont {Duan}},
  \bibinfo {author} {\bibfnamefont {W.-T.}\ \bibnamefont {Lyu}}, \bibinfo
  {author} {\bibfnamefont {G.-Y.}\ \bibnamefont {Wang}}, \bibinfo {author}
  {\bibfnamefont {J.-Y.}\ \bibnamefont {Zhu}}, \ and\ \bibinfo {author}
  {\bibfnamefont {E.}~\bibnamefont {Wang}},\ }\href {\doibase
  10.1140/epjc/s10052-024-13616-6} {\bibfield  {journal} {\bibinfo  {journal}
  {Eur. Phys. J. C}\ }\textbf {\bibinfo {volume} {84}},\ \bibinfo {pages}
  {1253} (\bibinfo {year} {2024})},\ \Eprint {http://arxiv.org/abs/2405.14235}
  {arXiv:2405.14235 [hep-ph]} \BibitemShut {NoStop}%
\bibitem [{\citenamefont {Dai}\ \emph {et~al.}(2026)\citenamefont {Dai},
  \citenamefont {Lyu},\ and\ \citenamefont {Oset}}]{Dai:2026zqn}%
  \BibitemOpen
  \bibfield  {author} {\bibinfo {author} {\bibfnamefont {L.~R.}\ \bibnamefont
  {Dai}}, \bibinfo {author} {\bibfnamefont {W.-T.}\ \bibnamefont {Lyu}}, \ and\
  \bibinfo {author} {\bibfnamefont {E.}~\bibnamefont {Oset}},\ }\href@noop {}
  {\bibfield  {journal} {\bibinfo  {journal} {Chinese Physics C}\ } (\bibinfo
  {year} {2026})},\ \bibinfo {note} {in press},\ \Eprint
  {http://arxiv.org/abs/2602.09136} {arXiv:2602.09136 [hep-ph]} \BibitemShut
  {NoStop}%
\bibitem [{\citenamefont {Ablikim}\ \emph {et~al.}(2022)\citenamefont {Ablikim}
  \emph {et~al.}}]{BESIII:2022cxi}%
  \BibitemOpen
  \bibfield  {author} {\bibinfo {author} {\bibfnamefont {M.}~\bibnamefont
  {Ablikim}} \emph {et~al.} (\bibinfo {collaboration} {BESIII}),\ }\href
  {\doibase 10.1103/PhysRevD.106.072006} {\bibfield  {journal} {\bibinfo
  {journal} {Phys. Rev. D}\ }\textbf {\bibinfo {volume} {106}},\ \bibinfo
  {pages} {072006} (\bibinfo {year} {2022})},\ \Eprint
  {http://arxiv.org/abs/2207.14350} {arXiv:2207.14350 [hep-ex]} \BibitemShut
  {NoStop}%
\bibitem [{\citenamefont {Bramon}\ \emph {et~al.}(1992)\citenamefont {Bramon},
  \citenamefont {Grau},\ and\ \citenamefont {Pancheri}}]{Bramon:1992kr}%
  \BibitemOpen
  \bibfield  {author} {\bibinfo {author} {\bibfnamefont {A.}~\bibnamefont
  {Bramon}}, \bibinfo {author} {\bibfnamefont {A.}~\bibnamefont {Grau}}, \ and\
  \bibinfo {author} {\bibfnamefont {G.}~\bibnamefont {Pancheri}},\ }\href
  {\doibase 10.1016/0370-2693(92)90041-2} {\bibfield  {journal} {\bibinfo
  {journal} {Phys. Lett. B}\ }\textbf {\bibinfo {volume} {283}},\ \bibinfo
  {pages} {416} (\bibinfo {year} {1992})}\BibitemShut {NoStop}%
\bibitem [{\citenamefont {Manohar}(1998)}]{Manohar:1998xv}%
  \BibitemOpen
  \bibfield  {author} {\bibinfo {author} {\bibfnamefont {A.~V.}\ \bibnamefont
  {Manohar}},\ }in\ \href@noop {} {\emph {\bibinfo {booktitle} {{Les Houches
  Summer School in Theoretical Physics, Session 68: Probing the Standard Model
  of Particle Interactions}}}}\ (\bibinfo {year} {1998})\ pp.\ \bibinfo {pages}
  {1091--1169},\ \Eprint {http://arxiv.org/abs/hep-ph/9802419}
  {arXiv:hep-ph/9802419} \BibitemShut {NoStop}%
\bibitem [{\citenamefont {Abreu}\ \emph {et~al.}(2023)\citenamefont {Abreu},
  \citenamefont {Dai},\ and\ \citenamefont {Oset}}]{Abreu:2023yvf}%
  \BibitemOpen
  \bibfield  {author} {\bibinfo {author} {\bibfnamefont {L.~M.}\ \bibnamefont
  {Abreu}}, \bibinfo {author} {\bibfnamefont {L.}~\bibnamefont {Dai}}, \ and\
  \bibinfo {author} {\bibfnamefont {E.}~\bibnamefont {Oset}},\ }\href {\doibase
  10.1016/j.physletb.2023.137999} {\bibfield  {journal} {\bibinfo  {journal}
  {Phys. Lett. B}\ }\textbf {\bibinfo {volume} {843}},\ \bibinfo {pages}
  {137999} (\bibinfo {year} {2023})},\ \Eprint
  {http://arxiv.org/abs/2303.00382} {arXiv:2303.00382 [hep-ph]} \BibitemShut
  {NoStop}%
\bibitem [{\citenamefont {He}\ \emph {et~al.}(2026)\citenamefont {He},
  \citenamefont {Liu}, \citenamefont {Geng}, \citenamefont {Guo},\ and\
  \citenamefont {Xie}}]{He:2026mkf}%
  \BibitemOpen
  \bibfield  {author} {\bibinfo {author} {\bibfnamefont {Y.-B.}\ \bibnamefont
  {He}}, \bibinfo {author} {\bibfnamefont {X.-H.}\ \bibnamefont {Liu}},
  \bibinfo {author} {\bibfnamefont {L.-S.}\ \bibnamefont {Geng}}, \bibinfo
  {author} {\bibfnamefont {F.-K.}\ \bibnamefont {Guo}}, \ and\ \bibinfo
  {author} {\bibfnamefont {J.-J.}\ \bibnamefont {Xie}},\ }\href {\doibase
  10.1103/5sr9-tzj6} {\bibfield  {journal} {\bibinfo  {journal} {Phys. Rev. D}\
  }\textbf {\bibinfo {volume} {113}},\ \bibinfo {pages} {L051501} (\bibinfo
  {year} {2026})},\ \Eprint {http://arxiv.org/abs/2407.13486}
  {arXiv:2407.13486} \BibitemShut {NoStop}%
\bibitem [{\citenamefont {Mandl}\ and\ \citenamefont {Shaw}(2010)}]{MandlShaw}%
  \BibitemOpen
  \bibfield  {author} {\bibinfo {author} {\bibfnamefont {F.}~\bibnamefont
  {Mandl}}\ and\ \bibinfo {author} {\bibfnamefont {G.}~\bibnamefont {Shaw}},\
  }\href@noop {} {\emph {\bibinfo {title} {Quantum Field Theory}}},\ \bibinfo
  {edition} {2nd}\ ed.\ (\bibinfo  {publisher} {John Wiley and Sons},\ \bibinfo
  {year} {2010})\BibitemShut {NoStop}%
\bibitem [{\citenamefont {Guo}()}]{Guo_private}%
  \BibitemOpen
  \bibfield  {author} {\bibinfo {author} {\bibfnamefont {F.~K.}\ \bibnamefont
  {Guo}},\ }\href@noop {} {}\bibinfo {note} {Private communication,
  (2026)}\BibitemShut {NoStop}%
\end{thebibliography}%

\end{document}